# AI-Driven Rapid Identification of Bacterial and Fungal Pathogens in Blood Smears of Septic Patients.

Agnieszka Sroka-Oleksiak[a], Adam Pardyl[b,c], Dawid Rymarczyk[b,d], Aldona Olechowska-Jarząb[e], Katarzyna Biegun-Drożdż[a], Dorota Ochońska[a], Michał Wronka[a], Adriana Borowa [d], Tomasz Gosiewski[f], Miłosz Adamczyk[b], Henryk Telega[b], Bartosz Zieliński[b], Monika Brzychczy-Włoch[a*]

[a] Department of Molecular Medical Microbiology, Chair of Microbiology, Faculty of Medicine, Jagiellonian University Medical College, Kraków, Poland
agnieszka.sroka@uj.edu.pl, k.biegun@uj.edu.pl , dorota.ochonska@uj.edu.pl, michal.wronka@alumni.uj.edu.pl, m.brzychczy-wloch@uj.edu.pl

[b] Faculty of Mathematics and Computer Science, Jagiellonian University Kraków, Poland
adam.pardyl@doctoral.uj.edu.pl, dawid.rymarczyk@uj.edu.pl, milosz2002.adamczyk@student.uj.edu.pl, henryk.telega@uj.edu.pl, bartosz.zielinski@uj.edu.pl

[c] Doctoral School of Exact and Natural Sciences, Jagiellonian University Kraków, Poland
adam.pardyl@doctoral.uj.edu.pl,

[d] Ardigen SA, Leona Henryka Sternbacha 1, 30-394 Kraków, Poland
ada.borowa@student.uj.edu.pl, dawid.rymarczyk@uj.edu.pl,

[e] Department of Microbiology, University Hospital, ul. Jakubowskiego 2, 30-688 Kraków, Poland. aldona.olechowska-jarzab@uj.edu.pl

[f] Microbiome Research Laboratory, Department of Molecular Medical Microbiology, Chair of Microbiology, Faculty of Medicine, Jagiellonian University Medical College, Kraków, Poland tomasz.gosiewski@uj.edu.pl

Corresponding authors:

1. Monika Brzychczy-Włoch

    Department of Molecular Medical Microbiology, Chair of Microbiology, Faculty of Medicine, Jagiellonian University Medical College, Czysta 18, 31-121 Krakow, Poland; m.brzychczy-wloch@uj.edu.pl

## Abstract

Sepsis is a life-threatening condition which requires rapid diagnosis and treatment. Traditional microbiological methods are time-consuming and expensive. In response to these challenges, deep learning algorithms were developed to identify 14 bacteria species and 3 yeast-like fungi from microscopic images of Gram-stained smears of positive blood samples from sepsis patients.

A total of 16,637 Gram-stained microscopic images were used in the study. The analysis used the Cellpose 3 model for segmentation and Attention-based Deep Multiple Instance Learning for classification. Our model achieved an accuracy of 77.15% for bacteria and 71.39% for fungi, with ROC AUC of 0.97 and 0.88, respectively. The highest values, reaching up to 96.2%, were obtained for *Cutibacterium acnes*, *Enterococcus faecium*, *Stenotrophomonas maltophilia* and *Nakaseomyces glabratus*. Classification difficulties were observed in closely related species, such as *Staphylococcus hominis* and *Staphylococcus haemolyticus*, due to morphological similarity, and within *Candida albicans* due to high morphotic diversity.

The study confirms the potential of our model for microbial classification, but it also indicates the need for further optimisation and expansion of the training data set. In the future, this technology could support microbial diagnosis, reducing diagnostic time and improving the effectiveness of sepsis treatment due to its simplicity and accessibility.

Part of the results presented in this publication was covered by a patent application at the European Patent Office EP24461637.1 "A computer implemented method for identifying a microorganism in a blood and a data processing system therefor".

## Highlights

1. A new deep learning-based method to support microbiological diagnostics of sepsis.
2. First study to use as many as 16637 microscopic images of pathogens in blood samples.
3. The identification of 14 bacteria in the blood (accuracy 77.15%, ROC AUC 0.97).
4. The identification of 3 fungi in the blood (accuracy 71.39%, ROC AUC 0.88).
5. The presented method is covered by EPO EP24461637.1.

## 1. Introduction

Sepsis is a life-threatening systemic inflammatory response of the body to the presence of microorganisms in the bloodstream, leading to organ damage and even organ failure [1]. It remains one of the most pressing medical challenges, with early diagnosis and initiation of the right treatment being essential. According to the WHO, there were 48.9 million cases of sepsis and 11 million associated deaths in 2020, accounting for 20% of all deaths worldwide [2]. Unlike other acute illnesses, sepsis is characterized by a wide range of symptoms, including fever and hypothermia, rapid breathing and heart rate, neurological deficits and skin rashes, making it difficult to develop standard therapies [3] and to make clinical decisions [4]. Therefore, several laboratory tests must be performed to confirm the diagnosis.

The conventional microbiology diagnosis of sepsis is based on the collection of blood from the patient, the culture of the pathogenic microorganism in a liquid medium culture in a closed automatic system, and the subsequent isolation and identification of the pathogenic microorganism using phenotypic or molecular tests (Figure 1A). The main limitation of culture-based methods is the long time taken to obtain results, which can range from several days to even several weeks [5]. Although molecular methods greatly reduce these times, they remain expensive and require dedicated laboratory facilities. As a result, broad-spectrum antibiotic therapy is often used, which is a major contributor to the spread of antibiotic resistance and the emergence of resistant pathogens.

Innovative diagnostic approaches are being explored. These include biomarkers such as chemokine ligand 7 (CXCL7) [6], presepsin, CD64 [7], damage-associated molecular patterns (DAMPs), non-coding RNA, microRNA [8], and others. The mentioned markers may indicate the development of sepsis, but they do not identify the pathogen, which is crucial for microbiological diagnostics. Moreover, most of these are of limited utility because they are applied to specific patient populations, such as CXCL7 in pediatric patients, or are limited by laboratory conditions and equipment (e.g. detection of microRNA and non-coding RNA) [9]. The complexity of the body's response to infection requires the combination of biomarker panels with patient-specific data (e.g. clinical, genetic, and demographic) to improve diagnostic accuracy. Consequently, there is an increasing need for more universal, inexpensive, simple and, most importantly, rapid methods to aid in the diagnosis of sepsis.

In response to the above limitations, this work focuses on the development of deep learning algorithms for the identification of selected species of bacteria and yeast-like fungi based on microscopic images of Gram-stained preparations from clinical materials - blood samples from sepsis patients. The primary motivation for our research was to shorten the diagnostic stages (Figure 1B), which would enable faster diagnosis and treatment while reducing the cost of diagnostics and increasing its accessibility.

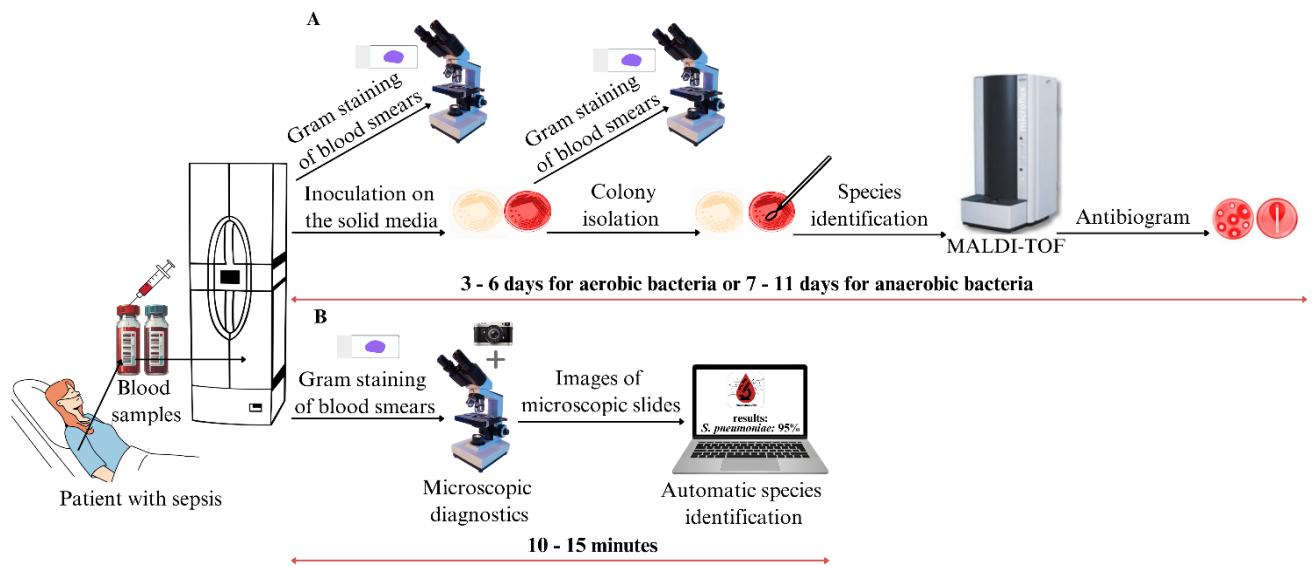

Figure 1. A) Standard microbiological diagnostics using culture, microscopy and mass spectrometry (MALDI-TOF) can last even several days. B) Our proposed method for microbiological diagnostics based on Gram-stained blood smear images analyzed with deep learning algorithms aims to shorten this time to several minutes.

Compared to other works, our approach is innovative because it focuses on detecting and classifying the pathogens responsible for the development of sepsis, while previous AI models focus mainly on predicting its occurrence. For example, left-aligned models predict the onset of sepsis from a specific point in time, such as a patient's admission to the hospital, with the aim of speeding up diagnosis and initiating treatment. In contrast, right-aligned models use large amounts of data to continuously predict the development of sepsis over time, allowing earlier diagnosis [7]. However, a major limitation of these methods is that they do not account for the identification of the microorganisms responsible for the infection, which significantly hinders the implementation of targeted therapy.

Our innovative approach overcomes these limitations by combining the power of artificial intelligence with classical microscopy to automatically and rapidly identify microorganisms at the species level. In the future, this solution could significantly shorten the diagnosis time and enable targeted antibiotic therapy at an early stage of treatment, thereby increasing the patient's chances of survival.

## 2. Materials and methods

### 2.1 Microbiological data base

The research has been approved by the the Research Ethics Committee of the Jagiellonian University Medical College (no. 118.0043.1.159.2024).
The study material was provided by the anonymized archival database of the Department of Microbiology at the University Hospital in Krakow. The images were obtained from blood smears of positive bottle culture from septic patients, stained by the Gram staining method using a PREVI®COLOR GRAM (bioMérieux, Marcy-l'Étoile, France) instrument and compatible with this machine's commercially available kit PREVI®COLOR GRAM (bioMérieux, Marcy-l'Étoile, France). Images were collected from 2 - 4 slides (per each positive blood sample) using an optical microscope with a 100x super apochromatic objective under oil immersion.

Patients were diagnosed as part of routine diagnostic procedures in the hospital according to current protocols. Anonymized results were provided for the purposes of this study. The identification of the bacterial species was based on a comprehensive diagnostic procedure. This consisted of an initial blood culture in liquid media: BACT/ALERT Culture Media Bottle (aerobic) FA; and BACT/ALERT Culture Media Bottle (anaerobic) FN (bioMérieux, Marcy-l'Étoile, France) in a closed system BACT/ALERT VIRTUO (bioMérieux, Marcy-l'Étoile, France). Subsequently, the culture was performed on solid media. Blood samples were inoculated on four media: Columbia CNA + 5% sheep blood (bioMérieux, Marcy-l'Étoile, France) for Gram-positive cocci, Mac Conkey Agar (bioMérieux, Marcy-l'Étoile, France) for Gram-negative rods, Schaedler + 5% sheep blood (bioMérieux, Marcy-l'Étoile, France) for anaerobic bacteria and Sabourad Gentamicin Chloramphenicol 2 (bioMérieux, Marcy-l'Étoile, France) for yeast-like fungi. The cultures were carried out for 18-48 hours under conditions appropriate for each type of microorganism, including a temperature range of 35-37°C for aerobic bacteria and a 5% $CO_2$ enriched atmosphere for anaerobic bacteria. Optimal conditions for fungi were 25°C and 48h of culture. After incubation, the bacterial and fungal colonies were subjected to species identification using MALDI-TOF (Bruker, Rheinstetten, Germany).

Based on a retrospective analysis conducted at the University Hospital in Krakow for the purposes of this study, the species of microorganisms most frequently identified as the etiological factors of sepsis in the years 2021–2023 were selected. Among them were: Gram-negative bacteria (*Escherichia coli*, *Klebsiella pneumoniae*, *Acinetobacter baumannii, Pseudomonas aeruginosa, Stenotrophomonas maltophilia),* Gram-positive bacteria (*Staphylococcus aureus*, *Staphylococcus epidermidis*, *Staphylococcus hominis*, *Staphylococcus haemolyticus*, *Streptococcus pneumoniae, Streptococcus mitis/oralis, Enterococcus faecium, Enterococcus faecalis*, *Cutibacterium acnes),* and yeast-like fungi *(Candida albicans, Candida parapsilosis, Nakaseomyces glabratus* - previously *Candida glabrata*, and *Kluyveromyces marxianus* – previously *Candida kefyr*).

The number of images used in the study for each species is presented in Table 1, while example images of microorganisms are shown in Figure 2.

Table 1. Quantitative summary of the number of patients, microscopic images, and cells of bacteria used in training and test sets. A number of images in training and test sets is an approximate average across all folds.

| | Species name | No. of patients | No. of microscopic images | No. of cells | No. of images in training set | No. of images in test set |
|---|---|---|---|---|---|---|
| Gram-negative | *Escherichia coli* | 27 | 1 676 | ~31 730 | ~1 117 | ~559 |
| | *Klebsiella pneumoniae* | 25 | 1458 | ~182 096 | ~972 | ~486 |
| | *Acinetobacter baumanii* | 11 | 550 | ~10 410 | ~367 | ~183 |
| | *Stenotrophomonas maltophilia* | 11 | 795 | ~61 192 | ~530 | ~365 |
| | *Pseudomonas aeruginosa* | 17 | 690 | ~27 388 | ~460 | ~230 |
| Gram-positive | *Staphylococcus aureus* | 17 | 1 011 | ~116 915 | ~697 | ~337 |
| | *Staphylococcus epidermidis* | 14 | 801 | ~109 115 | ~534 | ~267 |
| | *Staphylococcus hominis* | 9 | 762 | ~61 070 | ~508 | ~254 |
| | *Staphylococcus haemolyticus* | 8 | 575 | ~88 253 | ~381 | ~190 |
| | *Streptococcus pneumoniae* | 14 | 1 273 | ~100 772 | ~849 | ~424 |
| | *Streptococcus mitis/oralis* | 9 | 610 | ~220 089 | ~407 | ~203 |
| | *Enterococcus faecium* | 17 | 831 | ~40 456 | ~554 | ~277 |
| | *Enterococcus faecalis* | 13 | 671 | ~62 054 | ~447 | ~223 |
| | *Cutibacterium acnes* | 9 | 564 | ~43 367 | ~376 | ~188 |
| Other bacteria | | 47 | 2483 | ~262 846 | ~1 655 | ~827 |
| Yeast-like fungi | *Candida albicans* | 13 | 561 | ~7 712 | ~374 | ~187 |
| | *Candida parapsilosis* | 8 | 511 | ~6 606 | ~341 | ~171 |
| | *Nakaseomyces glabratus* | 10 | 457 | ~8 446 | ~307 | ~150 |
| Other fungi | | 7 | 358 | ~3922 | ~237 | ~117 |
| Total | | **286** | **16 637** | **~1 444 439** | **~11 113** | **~5 638** |

To make the method robust to misdiagnosis of etiologic factors for sepsis, images of other species of bacteria (n=27) and yeast-like fungi (n=2) were also used to ensure the deep learning model is aware that there can be other species than those 17 used in a study. Among 27 additional species were Gram-negative bacteria (*Achromobacter denitrificans, Bacteroides fragilis, Bacteroides thetaomicron, Bacteroides uniformis, Citrobacter braakii, Enterobacter cloacae, Morganella morganii, Pasteurella multocida, Raoutella* spp., *Salmonella enterica, Serratia marcescens)*, Gram-positive bacteria (*Arcanobacterium haemolyticum, Brevibacterium casei, Clostridium perfringens, Corynebacterium aurimucosum, Corynebacterium falsenii, Lactobacillus* spp., *Lactococcus* spp., *Listeria monocytogenes, Micrococcus luteus, Peptoniphilus asaccharolyticus, Streptococcus gallolyticus, Streptococcus pyogenes, Streptococcus agalactiae, Streptococcus dysagalactiae, Serratia marcescens)*, and yeast-like fungi (*Candida tropicalis* and *Kluyveromyces marxianus)*.

Gram-negative bacteria:

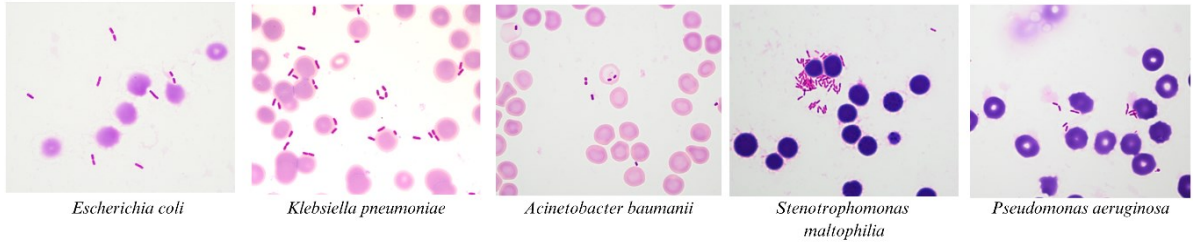

*Escherichia coli*   *Klebsiella pneumoniae*   *Acinetobacter baumanii*   *Stenotrophomonas maltophilia*   *Pseudomonas aeruginosa*

Gram-positive bacteria:

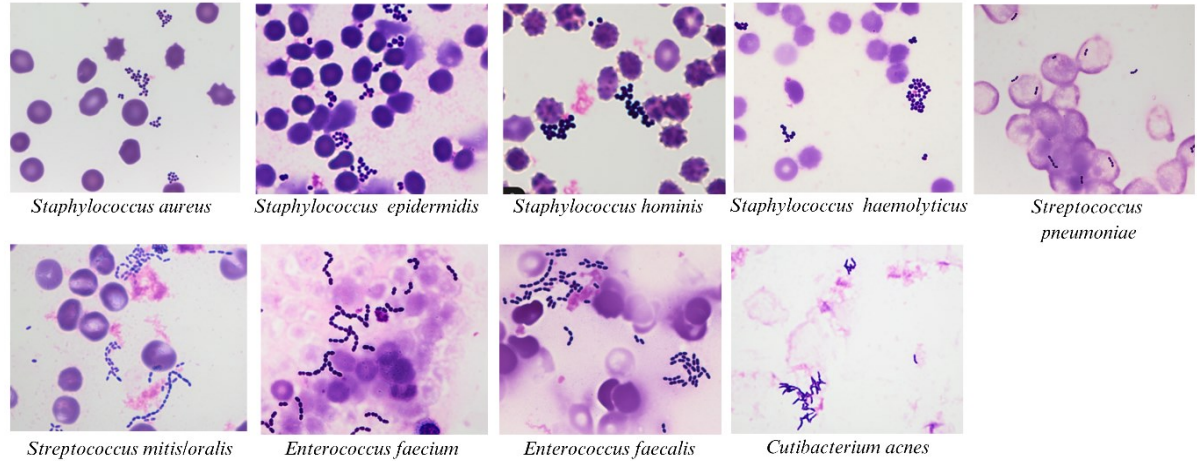

*Staphylococcus aureus*   *Staphylococcus epidermidis*   *Staphylococcus hominis*   *Staphylococcus haemolyticus*   *Streptococcus pneumoniae*

*Streptococcus mitis/oralis*   *Enterococcus faecium*   *Enterococcus faecalis*   *Cutibacterium acnes*

Yeast-like fungi:

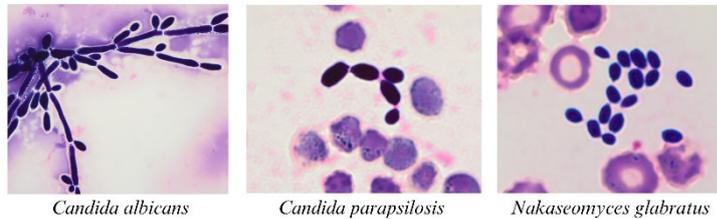

*Candida albicans*   *Candida parapsilosis*   *Nakaseomyces glabratus*

Figure 2. Example images of microscopic preparations of Gram-negative, Gram-positive bacteria and yeast-like fungi from the positive blood cultures of patients with sepsis from BACT/ALERT VIRTUO (bioMérieux, Marcy-l'Étoile, France) closed system. The slides were stained according to the Gram staining procedure using the PREVI®COLOR GRAM (bioMérieux, Marcy-l'Étoile, France) automatic instrument.

## 2.2. Pipeline of Deep learning-based classification of microorganisms.

To identify pathogens, we developed a pipeline (see Figures 3 and 4), which starts with segmentation of cells, cell-centered patch extraction, patches encoding, encodings aggregation and classification. For all experiments, we performed a stratified 3-fold cross validation (class-balanced), splitting the dataset according to the patient identifier into training and testing parts in 2:1 proportion (each patient is either in training or test set). We applied such split of data to prevent overfitting and reliably assess the effectiveness of the method.

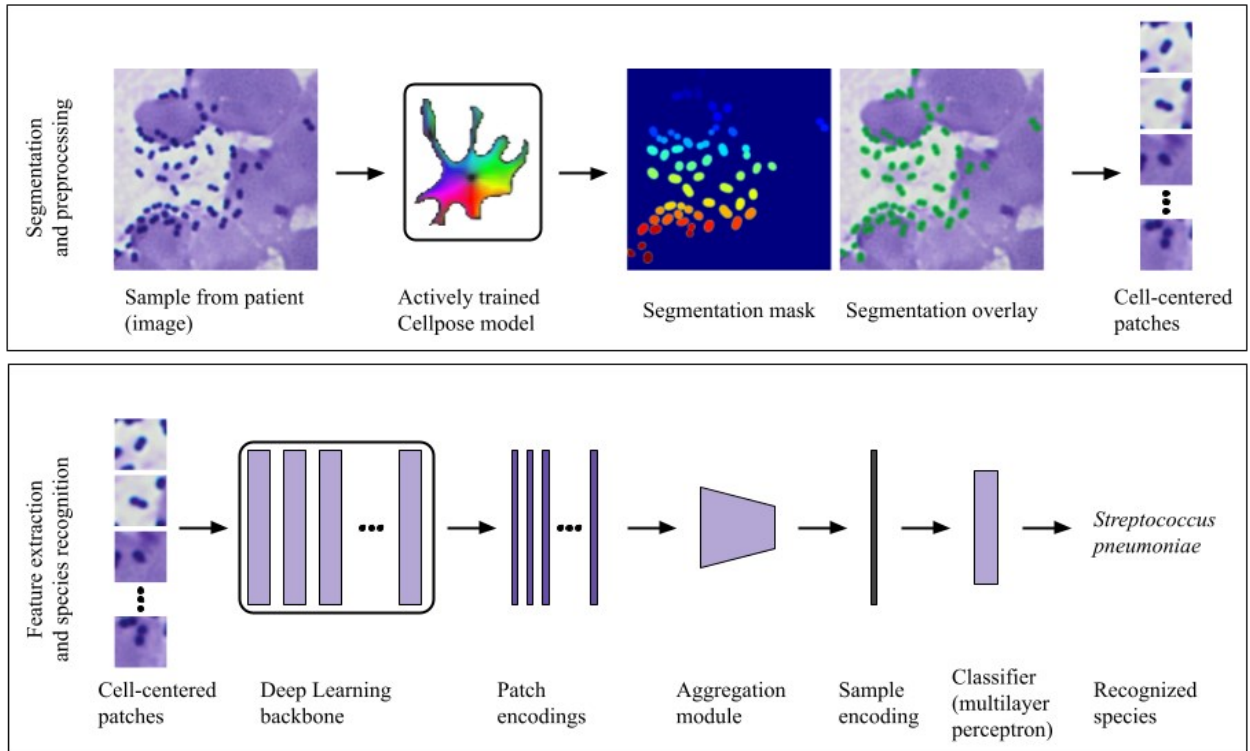

Figure 3. Overview of the decision-support systems recognizing species of microorganism from microscopic images. The image provides steps that the system is taking to perform a diagnosis, including segmentation of cells, cell-centered patch extraction, patches encoding by deep learning backbone, encodings aggregation and classification.

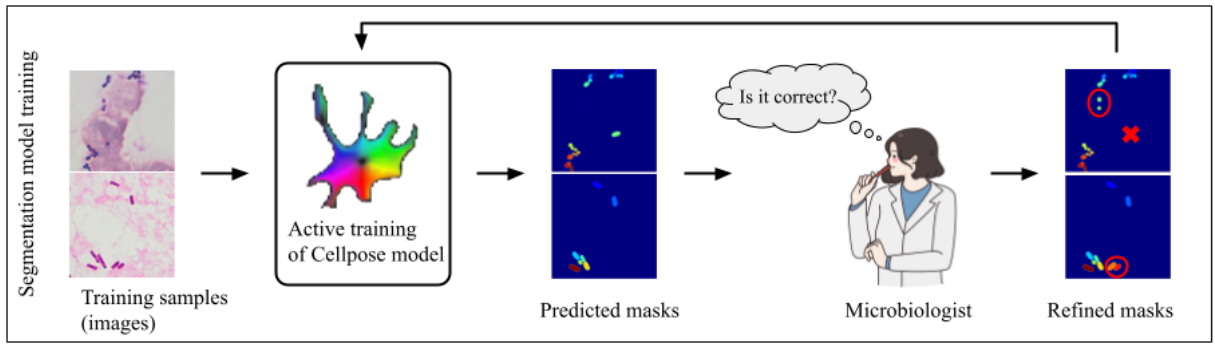

Figure 4. Overview of the active learning scenario. Cellpose model is corrected using microbiologist decisions about a presented segmentation.

## 2.3. Segmentation and extraction

We performed instance segmentation to separate bacteria from the visual background. Each bacterial cell was detected independently, and its location was passed to the next step of the pipeline. We used a fine-tuned Cellpose 3 [10], [11] convolutional neural network model to perform the segmentation. We divided the input images into 1024x1024 px patches and processed each of them separately, merging the resultant mask at the end. The fine-tuning process was performed with loss functions, parameters, and augmentations as described in [11]. Ground truth masks for fine-tuning consisted of 120 handcrafted segmentation masks for the bacteria and 656 for the yeast-like fungi, all manually prepared by microbiology experts (Figure 5).

Gram-negative bacteria: *Escherichia coli*

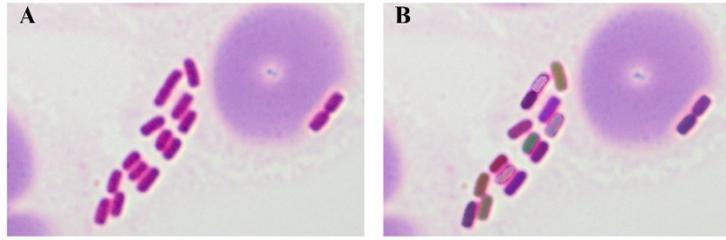

Gram-positive bacteria: *Enterococcus faecalis*

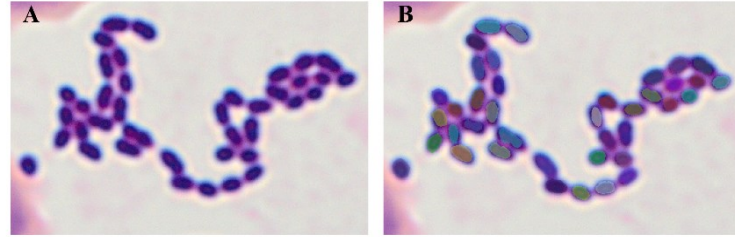

Yeast-like fungi: *Candida albicans*

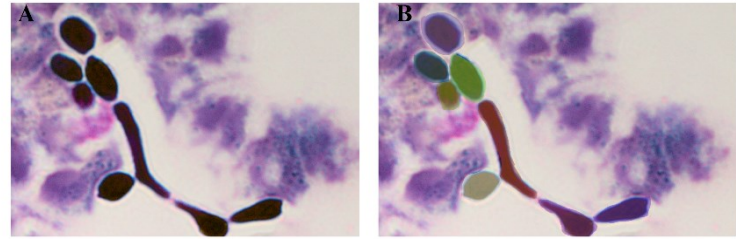

Figure 5. Sample image fragments of Gram-negative (*E. coli),* Gram-positive bacteria (*E. faecalis*), and yeast-like fungi (*C. albicans*) subjected to active learning. For each species, the original image fragment is shown on the left (A). On the right (B), we present the same fragment with an overlaid mask containing annotations from microbiologists.

We further extended the fine-tuning dataset using a semi-automated interactive ground truth creation process, with masks created by the pre-trained model, and manually checked by an expert, resulting in 1597 additional samples for the bacteria and 241 for the yeast-like fungi (Figure 6). The instance segmentation masks were used to extract regions of interest for further processing. For each segmented bacteria, we calculated its center and cropped out 96x96 px image fragment (patch) around it, representing a single bacteria and its surroundings.

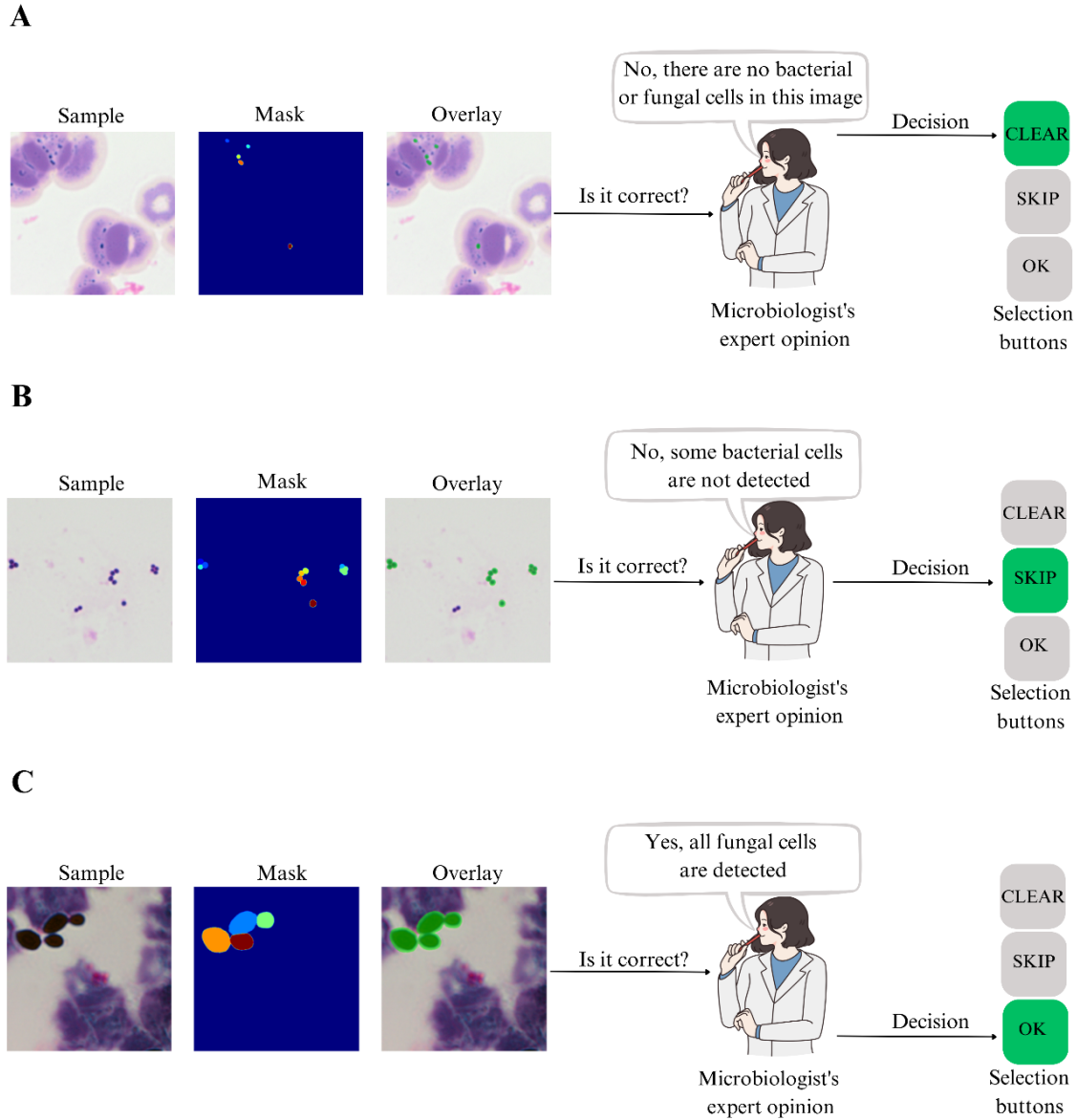

Figure 6. Sample images, corresponding masks of cells obtained automatically, and their overlay used in active learning. The user is asked to decide whether these masks are correctly predicted by the segmentation model or not. The "OK" button saves the predicted mask to be used for model training as ground truth, the "CLEAR" button replaces the current mask with an empty mask (to train Cellpose not to return any segmentations for this patch), and "SKIP" removes the patch (and the corresponding mask) from Cellpose training.

For the yeast-like fungi, we trained a separate model using the same pretrained Cellpose 3 and training setup. However, because the size of yeast-like cells is much more varied and often much larger than bacteria, we discarded all handcrafted cell masks with a diameter larger than 400 px (cell diameter is defined as the longest segment with both ends within the mask bounds). This was required to stabilise Cellpose training because, in the presence of the large variation in the fungi mask size, the model over-focused on the large outliers and underperformed on the common smaller cells. Furthermore, we used a much larger crop out of 800x800 px, to provide the model with a context of the cell surroundings. For training, all patches were downscaled to 224x224 pixels.

## 2.4. Bacteria classification

In this step, we computed neural representations of each extracted patch, combined the representations into a single vector describing the entire microscopic image, and finally classified it as one of the considered bacterium types (or the "other" class if none of the bacteria types matches). The patch representations were computed with a ResNet-50 [12] based encoder, pre-trained on the ImageNet dataset [13]. The resultant encoded representations of each patch extracted from one image were aggregated using the Attention-based Deep Multiple Instance Learning Pooling (AbMILP) [14] block, extended by adding multi-head attention as known from transformer networks [15]. The output was a single feature vector, which was finally classified by a single fully connected layer.

The encoder, attention pooling block, and final classifier were trained jointly as one neural network for 150 epochs with cross entropy as the loss function. We augmented the input patches with random horizontal and vertical flips, rotations, and crops, as well as random patch dropout. Additionally, we applied the Exponential Moving Average (EMA) to model weights to stabilise the training of the encoder. We used AdamW optimiser, with 0.01 weight decay, and a OneCycle learning rate scheduler [16] with a maximum rate of 3e-4 and 10 warmup epochs.

## 2.5. Fungi classification

We used the same algorithm for processing fungi images, with only a few minor changes. The patch representations were computed with an encoder based on a Small Visual Transformer [17] version of the DINO model [18], pre-trained on the ImageNet dataset. Notably, we did not train our encoder and relied on the DINO's ability to produce good representations even for previously unseen images. These resultant representations proved to be more reliable than encodings created by a specialist encoder, given the network will observe only around 1250 images during the training, too few to get a good understanding of its domain, resulting in overfitting of the encoder. AbMILP block and final classifier were trained jointly as one neural network for 45 epochs with the cross-entropy loss function. We used the same augmentation schema and optimizer settings as for bacteria images, but we did not apply EMA to the model weights.

## 3. Results and discussion

In this paper, we developed a deep learning-based system for species-level identification of selected bacteria and yeast-like fungi in clinical samples – blood smears of positive bottle culture from septic patients. To the best of our knowledge, based on the literature review and patent database analysis, this is the first study of its kind. Part of the results presented below was covered by a patent application at the European Patent Office EP24461637.1 "A computer implemented method for identifying a microorganism in a blood and a data processing system therefor".

## 3.1. Results for bacterial species identification without active learning

The average accuracy of our method without active learning for bacteria identification was 75.27%, with an average ROC AUC score of 0.97. Let us recall that the test set consisted of images that were not included in the training set. The analysis showed that the model achieved the highest level of accuracy in the classification of *C. acnes*, *S. maltophilia*, and *E.*

*faecium* (Figure 7). Slightly lower values were noted for *S. aureus S. epidermidis*, *E. faecalis*, and *S. pneumoniae*. On the other hand, *S. haemolyticus* and *A. baumanii* proved to be the most difficult to identify, with the lowest classification efficiency among the analyzed species.

| True Class \ Predicted Class | A. baumanii | C. acnes | E. faecalis | E. faecium | E. coli | K. pneumoniae | P. aeruginosa | S. aureus | S. epidermidis | S. haemolyticus | S. hominis | S. maltophilia | S. mitis oralis | S. pneumoniae | other |
|---|---|---|---|---|---|---|---|---|---|---|---|---|---|---|---|
| A. baumanii | 56.2 ±9.9 | 0.0 ±0.0 | 0.5 ±0.0 | 0.7 ±0.6 | 8.2 ±4.7 | 11.5 ±1.7 | 2.5 ±1.8 | 0.3 ±0.5 | 0.3 ±0.5 | 0.0 ±0.0 | 0.0 ±0.0 | 0.3 ±0.2 | 0.0 ±0.0 | 5.8 ±7.8 | 13.5 ±10.5 |
| C. acnes | 0.0 ±0.0 | 96.3 ±2.8 | 0.0 ±0.0 | 0.0 ±0.0 | 1.0 ±1.4 | 0.0 ±0.0 | 0.9 ±0.7 | 0.0 ±0.0 | 0.0 ±0.0 | 0.0 ±0.0 | 0.0 ±0.0 | 0.0 ±0.0 | 0.0 ±0.0 | 0.0 ±0.0 | 1.8 ±1.3 |
| E. faecalis | 0.2 ±0.2 | 0.2 ±0.2 | 81.6 ±13.7 | 1.9 ±2.3 | 0.7 ±0.2 | 0.0 ±0.0 | 0.5 ±0.4 | 2.5 ±3.5 | 0.2 ±0.2 | 0.0 ±0.0 | 0.1 ±0.2 | 0.2 ±0.2 | 3.2 ±3.0 | 5.7 ±4.1 | 3.2 ±3.0 |
| E. faecium | 0.8 ±0.7 | 0.0 ±0.0 | 1.7 ±2.3 | 91.3 ±5.0 | 2.9 ±3.2 | 0.0 ±0.0 | 0.1 ±0.2 | 1.2 ±1.7 | 0.3 ±0.4 | 0.0 ±0.0 | 0.3 ±0.4 | 0.1 ±0.2 | 0.1 ±0.2 | 0.1 ±0.1 | 1.0 ±0.4 |
| E. coli | 1.2 ±1.0 | 0.0 ±0.0 | 0.2 ±0.1 | 0.6 ±0.4 | 73.5 ±4.7 | 9.9 ±3.7 | 1.0 ±0.8 | 1.1 ±0.5 | 0.0 ±0.0 | 0.0 ±0.0 | 0.1 ±0.1 | 0.5 ±0.7 | 0.1 ±0.1 | 0.0 ±0.0 | 12.0 ±5.7 |
| K. pneumoniae | 1.7 ±1.5 | 0.0 ±0.0 | 0.1 ±0.1 | 0.1 ±0.1 | 17.1 ±5.4 | 75.0 ±9.5 | 0.3 ±0.4 | 0.7 ±0.5 | 0.0 ±0.0 | 0.0 ±0.0 | 0.0 ±0.0 | 0.1 ±0.1 | 0.0 ±0.0 | 0.0 ±0.0 | 4.8 ±3.6 |
| P. aeruginosa | 1.2 ±0.2 | 0.0 ±0.0 | 0.0 ±0.0 | 0.0 ±0.0 | 0.9 ±0.6 | 0.1 ±0.2 | 72.8 ±15.6 | 0.2 ±0.2 | 0.0 ±0.0 | 0.0 ±0.0 | 0.0 ±0.0 | 13.6 ±12.2 | 0.0 ±0.0 | 0.1 ±0.0 | 11.3 ±8.1 |
| S. aureus | 0.2 ±0.2 | 0.0 ±0.0 | 1.5 ±1.2 | 0.5 ±0.7 | 1.8 ±1.7 | 0.1 ±0.1 | 0.2 ±0.2 | 89.2 ±8.9 | 3.7 ±5.1 | 0.2 ±0.2 | 0.3 ±0.4 | 0.0 ±0.0 | 0.0 ±0.0 | 0.0 ±0.0 | 2.4 ±2.5 |
| S. epidermidis | 0.1 ±0.1 | 0.0 ±0.0 | 0.0 ±0.0 | 0.1 ±0.1 | 1.1 ±1.5 | 0.0 ±0.0 | 0.0 ±0.0 | 0.3 ±0.3 | 86.4 ±9.2 | 1.6 ±1.3 | 10.0 ±7.1 | 0.1 ±0.1 | 0.2 ±0.3 | 0.0 ±0.0 | 0.2 ±0.3 |
| S. haemolyticus | 0.0 ±0.0 | 0.0 ±0.0 | 0.0 ±0.0 | 0.0 ±0.0 | 0.6 ±0.8 | 0.0 ±0.0 | 0.0 ±0.0 | 1.2 ±1.8 | 9.3 ±6.2 | 45.6 ±12.6 | 41.5 ±18.6 | 0.0 ±0.0 | 0.0 ±0.0 | 0.0 ±0.0 | 1.8 ±2.5 |
| S. hominis | 0.0 ±0.0 | 0.0 ±0.0 | 0.0 ±0.0 | 0.3 ±0.2 | 0.1 ±0.2 | 0.0 ±0.0 | 0.0 ±0.0 | 0.0 ±0.0 | 11.0 ±13.7 | 21.8 ±18.5 | 66.6 ±13.7 | 0.0 ±0.0 | 0.0 ±0.0 | 0.0 ±0.0 | 0.2 ±0.3 |
| S. maltophilia | 0.0 ±0.0 | 0.0 ±0.0 | 0.0 ±0.0 | 0.0 ±0.0 | 0.9 ±1.2 | 0.0 ±0.0 | 2.8 ±2.3 | 0.0 ±0.0 | 0.0 ±0.0 | 0.0 ±0.0 | 0.0 ±0.0 | 92.7 ±4.1 | 0.0 ±0.0 | 0.0 ±0.0 | 3.6 ±4.1 |
| S. mitis oralis | 0.0 ±0.0 | 4.7 ±4.6 | 2.5 ±1.9 | 1.9 ±2.8 | 0.7 ±0.7 | 0.0 ±0.0 | 0.0 ±0.0 | 0.7 ±0.5 | 0.0 ±0.0 | 0.0 ±0.0 | 0.0 ±0.0 | 0.4 ±0.6 | 61.3 ±5.3 | 7.2 ±6.5 | 20.5 ±13.0 |
| S. pneumoniae | 0.0 ±0.0 | 0.0 ±0.0 | 5.2 ±3.5 | 0.0 ±0.0 | 0.8 ±1.2 | 0.2 ±0.1 | 0.0 ±0.0 | 0.0 ±0.0 | 0.0 ±0.0 | 0.0 ±0.0 | 0.0 ±0.0 | 0.0 ±0.0 | 0.3 ±0.4 | 83.5 ±8.8 | 10.0 ±6.9 |
| other | 1.5 ±1.6 | 1.0 ±0.3 | 1.4 ±0.3 | 3.6 ±4.6 | 10.6 ±3.5 | 1.5 ±1.0 | 4.7 ±0.2 | 0.2 ±2.0 | 0.5 ±0.7 | 1.3 ±1.7 | 1.2 ±1.0 | 2.6 ±3.1 | 6.5 ±5.4 | | 60.8 ±1.9 |

Figure 7. Averaged confusion matrix with standard deviation for bacterial species identification without active learning (the values are percentages).

### 3.2. Results for yeast-like fungi species identification without active learning

The average accuracy of our method without active learning for the identification of yeast-like fungi was 55.70%, and the average ROC AUC value was 0.80. Similarly to bacterial classification, the test set consisted of images that were not included in the training set. In the case of yeast-like fungi, the highest average accuracy was obtained for *N. glabratus* (66.7%) (Figure 8). The classification of the other species was less successful, especially for *C. albicans*, which had the lowest accuracy.

Figure 8. Averaged confusion matrix with standard deviation for yeast-like fungi identification without active learning (the values are percentages).

### 3.3. Results for bacterial species identification with active learning

As a result of introducing active learning into the study, the developed model for bacterial species identification achieved an accuracy of 77.15% (previously 75.27%) and an average ROC AUC score of 0.97. Higher performance was observed for 10 bacterial species. The greatest improvement was observed for *A. baumannii*, with prediction accuracy increased by more than 10% (from 56.2% to 66.6%), as well as for *S. pneumoniae* (from 83.5% to 87.6%) and S. *haemolyticus* (from 45.6% to 53.3%) - see Figure 9. On the other hand, for four bacteria species – *E. faecalis, S. hominis, K. pneumoniae* and *C. acnes* – a slight decrease in classification accuracy was observed, ranging from 0.1% to 2.4%. This slight drop in accuracy for individual species was due to the probabilistic nature of neural network training and did not significantly affect the overall effectiveness of the model.

Figure 9. Averaged confusion matrix with standard deviation for bacterial species identification with active learning (the values are percentages).

| True Class \ Predicted Class | A. baumanii | C. acnes | E. faecalis | E. faecium | E. coli | K. pneumoniae | P. aeruginosa | S. aureus | S. epidermidis | S. haemolyticus | S. hominis | S. maltophilia | S. mitis oralis | S. pneumoniae | other |
|---|---|---|---|---|---|---|---|---|---|---|---|---|---|---|---|
| A. baumanii | 66.6 ±13.3 | 0.0 ±0.0 | 0.3 ±0.2 | 0.5 ±0.4 | 1.6 ±0.9 | 7.7 ±1.1 | 0.2 ±0.2 | 0.0 ±0.0 | 0.3 ±0.5 | 0.0 ±0.0 | 0.0 ±0.0 | 0.3 ±0.5 | 0.0 ±0.0 | 5.7 ±6.8 | 16.7 ±9.4 |
| C. acnes | 0.0 ±0.0 | 96.2 ±2.7 | 0.0 ±0.0 | 0.0 ±0.0 | 0.5 ±0.7 | 0.0 ±0.0 | 1.3 ±1.0 | 0.0 ±0.0 | 0.0 ±0.0 | 0.0 ±0.0 | 0.0 ±0.0 | 0.2 ±0.2 | 0.0 ±0.0 | 0.0 ±0.0 | 1.8 ±1.4 |
| E. faecalis | 0.0 ±0.0 | 0.0 ±0.0 | 79.2 ±11.4 | 4.2 ±5.9 | 0.0 ±0.0 | 0.2 ±0.2 | 0.0 ±0.0 | 1.5 ±2.1 | 0.8 ±1.2 | 0.0 ±0.0 | 0.0 ±0.0 | 0.0 ±0.0 | 4.8 ±4.3 | 4.0 ±3.3 | 5.3 ±4.8 |
| E. faecium | 0.0 ±0.0 | 0.0 ±0.0 | 2.6 ±1.9 | 93.3 ±5.0 | 0.1 ±0.2 | 0.0 ±0.0 | 0.0 ±0.0 | 0.0 ±0.0 | 0.1 ±0.2 | 0.0 ±0.0 | 0.3 ±0.4 | 0.0 ±0.0 | 0.0 ±0.0 | 0.1 ±0.2 | 3.4 ±3.6 |
| E. coli | 0.9 ±1.2 | 0.0 ±0.0 | 0.0 ±0.0 | 0.0 ±0.0 | 74.7 ±10.8 | 10.1 ±7.2 | 1.5 ±1.3 | 0.0 ±0.0 | 0.0 ±0.0 | 0.0 ±0.0 | 0.0 ±0.0 | 0.3 ±0.4 | 0.0 ±0.0 | 0.0 ±0.0 | 12.5 ±7.6 |
| K. pneumoniae | 2.0 ±1.8 | 0.0 ±0.0 | 0.0 ±0.0 | 0.2 ±0.3 | 17.2 ±4.1 | 72.7 ±7.8 | 0.0 ±0.0 | 0.0 ±0.0 | 0.0 ±0.0 | 0.0 ±0.0 | 0.0 ±0.0 | 0.0 ±0.0 | 0.0 ±0.0 | 0.0 ±0.0 | 7.9 ±4.7 |
| P. aeruginosa | 0.0 ±0.0 | 0.1 ±0.2 | 0.2 ±0.3 | 0.0 ±0.0 | 0.5 ±0.5 | 0.0 ±0.0 | 73.8 ±15.7 | 0.1 ±0.2 | 0.0 ±0.0 | 0.0 ±0.0 | 0.0 ±0.0 | 16.6 ±18.0 | 0.0 ±0.0 | 0.0 ±0.0 | 8.6 ±5.8 |
| S. aureus | 0.0 ±0.0 | 0.0 ±0.0 | 0.7 ±0.6 | 0.0 ±0.0 | 0.1 ±0.1 | 0.0 ±0.0 | 0.0 ±0.0 | 91.9 ±6.4 | 4.1 ±5.4 | 1.1 ±1.5 | 0.5 ±0.1 | 0.0 ±0.0 | 0.0 ±0.0 | 0.0 ±0.0 | 1.5 ±1.3 |
| S. epidermidis | 0.0 ±0.0 | 0.0 ±0.0 | 0.0 ±0.0 | 0.2 ±0.3 | 0.0 ±0.0 | 0.0 ±0.0 | 0.0 ±0.0 | 0.2 ±0.3 | 88.1 ±10.1 | 1.0 ±0.8 | 10.3 ±9.6 | 0.0 ±0.0 | 0.0 ±0.0 | 0.0 ±0.0 | 0.1 ±0.1 |
| S. haemolyticus | 0.0 ±0.0 | 0.0 ±0.0 | 0.0 ±0.0 | 0.2 ±0.3 | 0.2 ±0.3 | 0.0 ±0.0 | 0.0 ±0.0 | 0.2 ±0.3 | 9.3 ±8.5 | 53.3 ±11.9 | 35.5 ±20.0 | 0.0 ±0.0 | 0.0 ±0.0 | 0.0 ±0.0 | 1.3 ±1.9 |
| S. hominis | 0.0 ±0.0 | 0.0 ±0.0 | 0.0 ±0.0 | 0.3 ±0.4 | 0.0 ±0.0 | 0.0 ±0.0 | 0.0 ±0.0 | 0.0 ±0.0 | 10.0 ±13.4 | 24.8 ±19.1 | 64.6 ±12.8 | 0.0 ±0.0 | 0.0 ±0.0 | 0.0 ±0.0 | 0.2 ±0.3 |
| S. maltophilia | 0.1 ±0.2 | 0.0 ±0.0 | 0.0 ±0.0 | 0.0 ±0.0 | 0.1 ±0.2 | 0.0 ±0.0 | 2.1 ±0.5 | 0.0 ±0.0 | 0.0 ±0.0 | 0.0 ±0.0 | 0.1 ±0.2 | 93.3 ±2.6 | 0.0 ±0.0 | 0.0 ±0.0 | 4.2 ±2.2 |
| S. mitis oralis | 0.0 ±0.0 | 7.8 ±8.8 | 2.4 ±2.0 | 1.2 ±1.7 | 0.0 ±0.0 | 0.0 ±0.0 | 0.0 ±0.0 | 0.4 ±0.6 | 0.0 ±0.0 | 0.0 ±0.0 | 0.0 ±0.0 | 0.0 ±0.0 | 62.7 ±6.6 | 10.5 ±5.9 | 15.0 ±4.2 |
| S. pneumoniae | 0.1 ±0.1 | 0.0 ±0.0 | 3.1 ±3.4 | 0.4 ±0.4 | 0.3 ±0.0 | 0.0 ±0.0 | 0.0 ±0.0 | 0.0 ±0.0 | 0.0 ±0.0 | 0.0 ±0.0 | 0.0 ±0.0 | 0.0 ±0.0 | 0.2 ±0.1 | 87.6 ±11.1 | 8.4 ±9.3 |
| other | 1.7 ±1.9 | 1.9 ±1.0 | 2.3 ±0.6 | 4.1 ±5.3 | 10.8 ±1.1 | 1.6 ±1.2 | 4.1 ±3.2 | 0.1 ±0.1 | 0.4 ±0.3 | 0.6 ±0.8 | 2.2 ±3.1 | 1.7 ±1.6 | 3.3 ±3.5 | 6.2 ±4.6 | 59.0 ±3.7 |

Even after applying active learning, some of the species were still difficult to classify, such as *S. haemolyticus* and *S. hominis*, which are closely related genetically and also morphologically similar, challenging even experienced diagnosticians. These species also had the highest misclassification rates, and they were most often confused with each other or with *S. epidermidis*. Similar difficulties were observed with *Streptococcus* genus, where *S. mitis/oralis* was classified as *S. pneumoniae*, and among the Gram-negative rods, where *E. coli* and *K. pneumoniae* were misidentified. In the case of *E. coli* and *K. pneumoniae*, it turned out that although we had the largest collection of images for these species compared to the other bacteria, our model achieved classification accuracies of 74.7% and 72.7%, respectively, suggesting other potential causes of problems in distinguishing between these species. One is undoubtedly the great similarity in morphology between these two species. Another reason may be the variation in cell size and length within the species, which was probably related to the different times of incubation of blood samples in a closed system and thus the lower or higher availability of nutrients. For example, *E. coli* showed high morphological variation, covering a wide range of phenotypes from short coccobacilli-like forms to elongated cells. The slightly lower accuracy of *K. pneumoniae* classification compared to *E. coli* may also be due to the presence of the different thicknesses of the polysaccharide bacterial capsule. The presence of the capsule is a feature that determines the different appearance of stained *K. pneumoniae* cells, which can make them difficult to identify correctly. This was also confirmed by previous research [19]. As in the case of *E. coli*, the difficulties in classifying *A. baumannii* (66.6%) were due to the wide morphological diversity of this species, which can also take the form of a bacilli or cocci. Detailed results obtained for each fold are provided in the Supplementary Appendix (Appendix A).

The "*Other*" category achieved moderate performance of 59.0%. The result indicates the difficulty of the model in accurately assigning samples to this category. This was probably due to the small number of images for bacterial species in this group, the high diversity of microorganisms, and their high morphological similarity to the considered bacteria.

The analysis of the ROC plots (Appendix B and C) showed that our model has a high classification quality. The ROC curves obtained for each of the three cross-validation folds were stable and close to each other. This indicates a high repeatability of the results regardless of the partitioning of the training data. This was also confirmed by the average area under the curve (AUC) values, especially for the following species among Gram-negative bacteria: *S. maltophilia* (mean AUC = 0.99), *K. pneumoniae* (mean AUC = 0.94), *A. baumannii* (mean AUC = 0.93), *E. coli* (mean AUC = 0.92), (Appendix B) and among Gram-positive bacteria: *S. aureus*, *C. acnes*, *S. epidermidis*, *S. pneumoniae*, and *E. faecium* for which AUC was ≥ 0.98 - see Appendix C.

*3.4. Results for yeast-like fungi species identification*

After application of active learning, the developed model for identification of yeast-like fungal species achieved an accuracy of 71.4% and a ROC AUC score of 0.88. The best and statistically significant results were reached for the *N. glabratus* (from 66.7% to 93.5%), *C. parapsilosis* (from 56.4% to 66%) and also for the "other" category (from 59.5% to 83.7%) (Figure 10 A). Incorrect predictions for *N. glabratus* class represented a small percentage, i.e. the model assigned 3.0% of the samples to *C. albicans*, 3.3% to *C. parapsilosis* and 0.2% to the 'other' class. A possible explanation for this observation is that *N. glabratus* has relatively small blastospores compared to other yeast-like fungi. Considering that the developed algorithms were also trained to classify bacterial species, our model was able to recognise the blastospores of *N. glabratus* as bacterial cells with a similar appearance, especially in terms of size and shape. The poorest results were observed for *C. albicans* (from 40.2% to 42.3%). Despite the fact that this species is characteristic for microbiologists to distinguish from other fungi in microscopic images, the model had clear difficulties in classifying it. Undoubtedly, its varied morphology contributed to this, as *Candida albicans* can appear both as single yeast-like cells called blastospore (characteristic of the other fungal species studied) and as long and often branched pseudohyphae and hyphae, which can also stain with varying intensity, increasing the complexity and difficulty in the segmentation stage of our method. Moreover, the model had difficulty distinguishing this class from *C. parapsilosis,* probably due to the similar morphology of their blastospores. Detailed results obtained for each fold are provided in the Supplementary Appendix (Appendix D).

Figure 10. Averaged confusion matrix with standard deviation for identifying yeast-like fungi species with active learning (the values are percentages).

The analysis of the ROC plots confirmed the above results and showed that our model has a high quality of classification of yeast-like fungi only with respect to *N. glabratus* (mean AUC = 0.99). For the other species, *C. parapsilosis* and *C. albicans*, the ability to discriminate between species was moderate, estimated at mean AUC = 0.82 and mean AUC = 0.77, respectively (Appendix E). We hypothesize that the classification of fungal species can be further improved by introducing a fungal-specific segmentation module. Unlike Cellpose, it should be able to detect shapes other than spheroids. We plan to address this issue in future research.

### 3.5. Distinguishing species within and across genera

According to the confusion matrix, the highest number of misclassifications was observed for Gram-positive bacteria within the genus *Staphylococcus*. This raised the question whether distinguishing between species of the same genus was more problematic for the model than distinguishing between species of different genera. To answer this question, we performed additional experiments, each dedicated to distinguishing four species within and across genera.

**Species across genera.** Firstly, we considered representative Gram-positive bacterial species with different morphology and cell arrangement: *C. acnes* (bacilli), *S. aureus* (staphylococci), *E. faecalis* (streptococci), and *S. pneumoniae* (diplococci).

The achieved accuracy of 95.7% and ROC AUC of 0.996 showed that the model classifies most species very well, especially *C. acnes* and *S. aureus*, with correct classifications above 98% (Figure 11A). Slightly worse classification results were observed for *E. faecalis*, recognised by the model as *S. pneumoniae* species. The obtained results were clearly reflected in the t-SNE analysis graph (Figure 11B), which showed separate clusters belonging to *C. acnes* and *S. aureus* and overlapping clusters of *E. faecalis* and *S. pneumoniae*. Both *E. faecalis* and *S. pneumoniae* are streptococci, and it is difficult for the human eye to distinguish between their morphology and the arrangement of their cells in a microscopic image. Despite these challenges, the performance of the model is very good, highlighting its ability to classify species even with morphologies that are difficult to distinguish.

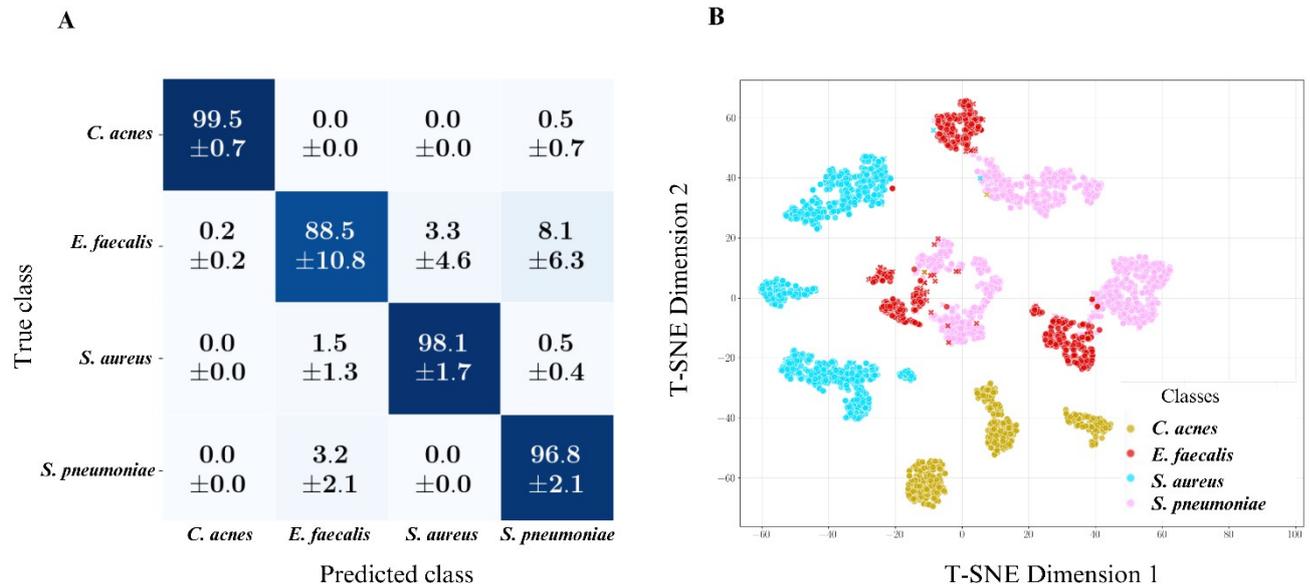

Figure 11. **A)** Averaged confusion matrix with standard deviation for distinguishing species from four different genera (*C. acnes, E. faecalis, S. aureus, S. pneumoniae*) of the group of Gram-positive bacteria (the values are percentages), and **B)** T-SNE plot showing the spatial distribution of individual bacterial species. The model distinguishes well between the bacteria species of different morphology.

**Species within genera.** Secondly, we considered Gram-positive bacterial species with similar morphology and cell arrangements belonging to *Staphylococcus* genus: *S. aureus*, *S. epidermidis, S. haemolyticus,* and *S. hominis.*

The assumptions regarding the difficulty of distinguishing species within the same genus were confirmed by the obtained accuracy of 72.30% with ROC AUC of 90.32%. However, the developed model reached a high efficiency in the recognition of the species *S. aureus*, with a correct classification rate of 93.4% (Figure 12A).

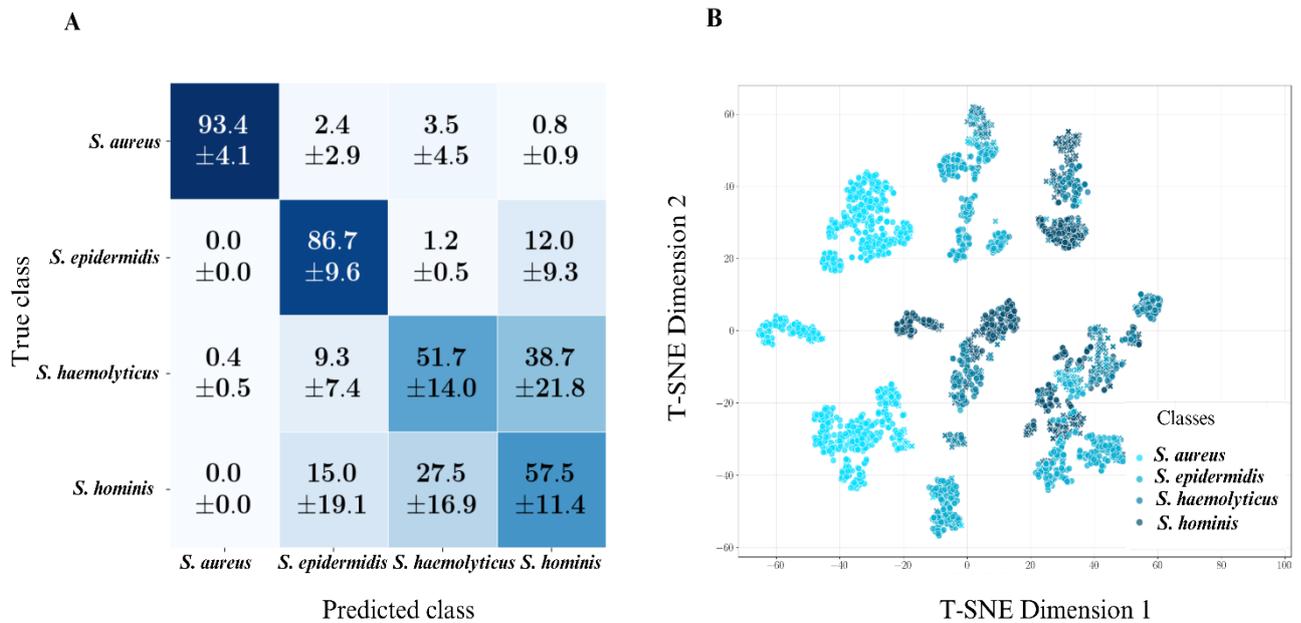

Figure 12. **A)** Averaged confusion matrix with standard deviation for distinguishing four different species (*S. aureus, S. epidermidis, S. haemolyticus, S. hominis*) within one genus *Staphyloccocus* spp. (the values are percentages), and **B)** T-SNE plot showing the spatial distribution of individual bacterial species. The model struggles when detecting bacteria species of similar morphological profiles. However, the results are still of high fidelity.

The accuracy of *S. epidermidis* classification was relatively high (86.7%), but the most common mistake was to confuse it with *S. hominis*. The model had the most difficulty distinguishing between *S. haemolyticus* and *S. hominis*, which had the lowest classification accuracy. In particular, *S. haemolyticus* was often misidentified as *S. hominis* or *S. epidermidis,* and *S. hominis* as *S. haemolyticus.* An additional factor that may have influenced the lower classification accuracy of these species was the much smaller number of their images in the training dataset, which may have limited the model's ability to learn their unique characteristics. The distribution of points on the t-SNE plot (Figure 12B), representing the number of patients, confirmed that the *S. aureus* species form well-separated clusters, whereas *S. epidermidis*, *S. haemolyticus,* and *S. hominis* overlapped, indicating smaller differences between these species compared to *S. aureus.*

### 3.4. Related works

So far, we have identified only four previous works that have classified bacteria in bloodstream infections. The work of Smith et al. (2018) analysed Gram-stained microscopic images of positive blood cultures [20]. However, they were classified only into four categories: Gram-positive cocci in pairs or chains, Gram-positive cocci in clusters, Gram-negative bacilli, or background. Those categories are much broader than our approach, which is much more relevant from the clinical practice perspective.

Kim et al. [21] proposed a method to identify four bacteria types: *Escherichia, Staphylococcus, Enterococcus, Streptococcus,* and the "other" class, which included less common bacteria. However, this work introduces only a protocol for analysing such data. In the following study of Kim et al. [22], they analysed the accuracy in such a scenario, achieving around 90% accuracy in classifying Gram-positive versus Gram-negative species. Another study by Iida et al. [23] classified bacteria not only in blood but also in other clinical materials such as sputum, faeces, pus, and urine. They also based their classification, as in the work of Smith et al., on staining and shape: Gram-positive cocci, Gram-positive bacilli, Gram-negative cocci, Gram-negative bacilli, as well as the categories Gram-positive unknown and Gram-negative unknown. Similarly to our approach, they first performed segmentation of bacteria cells (with Canny filter), and then applied a classifier, either SVM (Support Vector Machine) or DNN (Deep Neural Network). For blood samples, the models showed moderate accuracy in staining classification (around 59-60%) and slightly higher accuracy in recognising bacterial shapes, especially for cocci (up to 77% using SVM). SVM proved to be faster, but DNN had greater potential in analysing complex images.

A higher bacterial classification accuracy compared to Iida et al. was achieved in a similar study by Lejon and Andersson [24]. They firstly detected red blood cells to eliminate them from the analysis based on Dempster-Shafer and Fourier descriptors. Subsequently, they performed bacteria species classification based on the cells segmented by the Otsu algorithm and described via color descriptors. As a result, they obtained 89% accuracy for the classification of bacteria into Gram-positive and Gram-negative species.

The above literature data clearly shows that it is difficult to accurately identify genus or species of microorganisms in clinical material, including blood, even using advanced deep learning methods. As a result, most research is limited to classifying bacteria based on their shape and colour. While this data is important in the clinical diagnostic process, it does not provide information beyond what a laboratory diagnostician can obtain from standard microscopic observation.

Another effective and quick approach to identifying bacterial species and, in some cases, determining their antibiotic resistance is Surface Enhanced Raman Scattering (SERS) combined with DNN analysis. One study using this method with DNN achieved a detection accuracy of 98.68% for pathogenic bacteria (*Escherichia coli, Klebsiella pneumoniae, Acinetobacter baumannii, Enterococcus faecium, Enterococcus faecalis, Staphylococcus aureus* and *Pseudomonas aeruginosa*) and an accuracy of 99.85% for identifying carbapenem-resistant *Klebsiella pneumoniae* [25]. The potential of SERS with DNN goes well beyond the ability to identify microorganisms using Gram-stained microscopic images. However, in the context of potential application in clinical microbiology diagnostics, the method has several limitations, including the need to purchase a high-tech Raman microscope and nanoparticles, as well as expertise, which significantly increases the cost of implementation. Blood samples require careful preparation prior to analysis, including differential sedimentation or lysis of erythrocytes, and the presence of other biomolecules, such as haemoglobin, can negatively affect the quality of the results. In addition, instrument calibration requires a high degree of accuracy and Raman spectra acquisition takes a long time. Therefore, each step involves additional time and the possibility of technical error, which limits the scalability of this method in clinical diagnostics. In comparison to SERS, our solution is based on a very simple, cheap and, above all, laboratory-available method of staining and imaging.

## 3.5. Strengths and limitations of our study

The results obtained by our team indicate that bacterial and fungal identification at the species level from blood samples with use of developed method is possible with the high efficiency (ROC AUC values of 0.97 for bacteria and 0.88 for fungi). These values are very satisfactory, especially considering that the study included clinical samples – positive blood culture smears, which are a heterogeneous material containing numerous morphotic elements (e.g. white blood cells, red blood cells) in addition to bacteria and yeast-like fungi, as well as contaminants, and show significant differences in quality between samples, making their analysis even more difficult. Moreover, the study included many species that can be assigned to a single morphological type, such as Gram-negative bacilli, Gram-positive staphylococci, Gram-positive streptococci, and Gram-positive bacilli.

The first important factor that made it possible to obtain presented results is an appropriately prepared database of microscopic images. Most of the available online databases contain single images of Gram-stained slides. These are often of low resolution and usually lack a defined magnification. Others contain images that have been graphically corrected, which limits their usefulness for machine learning applications. Furthermore, none of them, unlike our database, contains microscopic images of clinical materials obtained at the appropriate magnification. In our research, the images were obtained at 100x magnification, compared to 40x in other studies, where such magnification makes it impossible to capture the subtle differences in cell morphology that are crucial for species classification. Additionally, our database consisted of as many as 16,637 Gram-stained images from blood smears, which were carefully selected to exclude those that were out of focus, of poor resolution, or contained artefacts. This allowed the model to be trained on high quality data, which was reflected in its ability to discriminate between different species and reduce the number of false predictions. Other important aspect of our database is its diversity in terms of the number of species and their morphology, which allowed the model to learn a wide range of features. Another component for achieving high accuracy was the combination of active learning, which optimized the selection of cell-centred training patches, and a multiple instance learning approach with an aggregation module, which robustly summarized cells detected in the image.

Although the results presented in this work are considered very satisfactory at this stage of the research, they indicate the need for further improvement of the model. One limitation is the insufficient number of images for some species, which made it difficult for the model to learn to identify them. This was particularly evident in the case of yeast-like fungi, for which we achieved lower accuracy than bacteria. There is also a need to include more examples with different morphological forms in the dataset, especially for species such as *Candida albicans.* In addition, our images of microorganisms were taken with a single microscope and camera, which may reduce the classification accuracy when the model is applied to images taken with lower quality equipment. Additionally, we have not investigated how the model reacts to samples from different facility and how batch effects are affecting the model performance. Finally, the method does not have any mechanism to quantify the uncertainty and reliability of the prediction, which we plan to address in future research.

## 4. Conclusion

In this paper, we propose a deep learning model to identify 14 species of bacteria and 3 species of fungi based on Gram-stained microscopic images obtained from positive blood cultures of patients with sepsis. This is the first study to attempt to identify microorganisms in blood at the species level.

Detailed analysis showed that the use of active learning with the participation of a microbiology expert significantly improved the results of bacterial and fungal classification, which plays an important role in the process of optimising artificial intelligence algorithms. In addition, the use of high-resolution microscopic images, carefully selected for their diagnostic quality, was an important factor in the high effectiveness of the model.

The results obtained for both bacteria (accuracy 77.15% and average ROC AUC score 0.97) and fungi (accuracy 71.39% and ROC AUC score 0.88) provide a solid basis for further development of the model. In order to increase the accuracy and to extend the classification to new microorganism species, it is necessary to perform similar studies in the future, but on larger and more diverse image sets.

## CRediT authorship contribution statement

**Agnieszka Sroka-Oleksiak:** Writing – original draft, Formal analysis, Visualization, **Adam Pardyl:** Writing – original draft, Methodology, Software, Validation, **Dawid Rymarczyk:** Writing – review and editing, Formal analysis, Conceptualization, **Aldona Olechowska-Jarząb:** Resources, Writing – review and editing, **Katarzyna Biegun-Drożdż:** Investigation, Writing-review and editing, **Dorota Ochońska:** – Validation, Writing-review and editing, **Michał Wronka:** Formal analysis, Software, Writing – review and editing, **Adriana Borowa:** Writing – review and editing, Visualization, **Tomasz Gosiewski:** Writing – review and editing, **Miłosz Adamczyk:** Software, **Henryk Telega** – Data curation, **Bartosz Zieliński:** Methodology, Validation, Formal analysis, Writing – review & editing, Supervision, Conceptualization, **Monika Brzychczy-Włoch:** Conceptualization, Methodology, Validation, Formal analysis, Writing – review & editing, Supervision, Project administration, Funding acquisition,

## Declaration of competing interest

The authors declare that they have no known competing financial interests or personal relationships that could have appeared to influence the work reported in this paper.

## Acknowledgements

This work was supported by the Foundation for Polish Science co-financed by the European Union under the European Funds for Smart Economy 2021-2027 (grant number: FENG.02.07-IP.05-0068/23).

The authors would like to thank the management of the University Hospital in Krakow for giving their permission to access the archival collection of anonymized microscopic images from blood smears of sepsis patients, which were used in this research.

## Data availability

Data will be made available on request.

**Supplementary data**

**Appendix A**

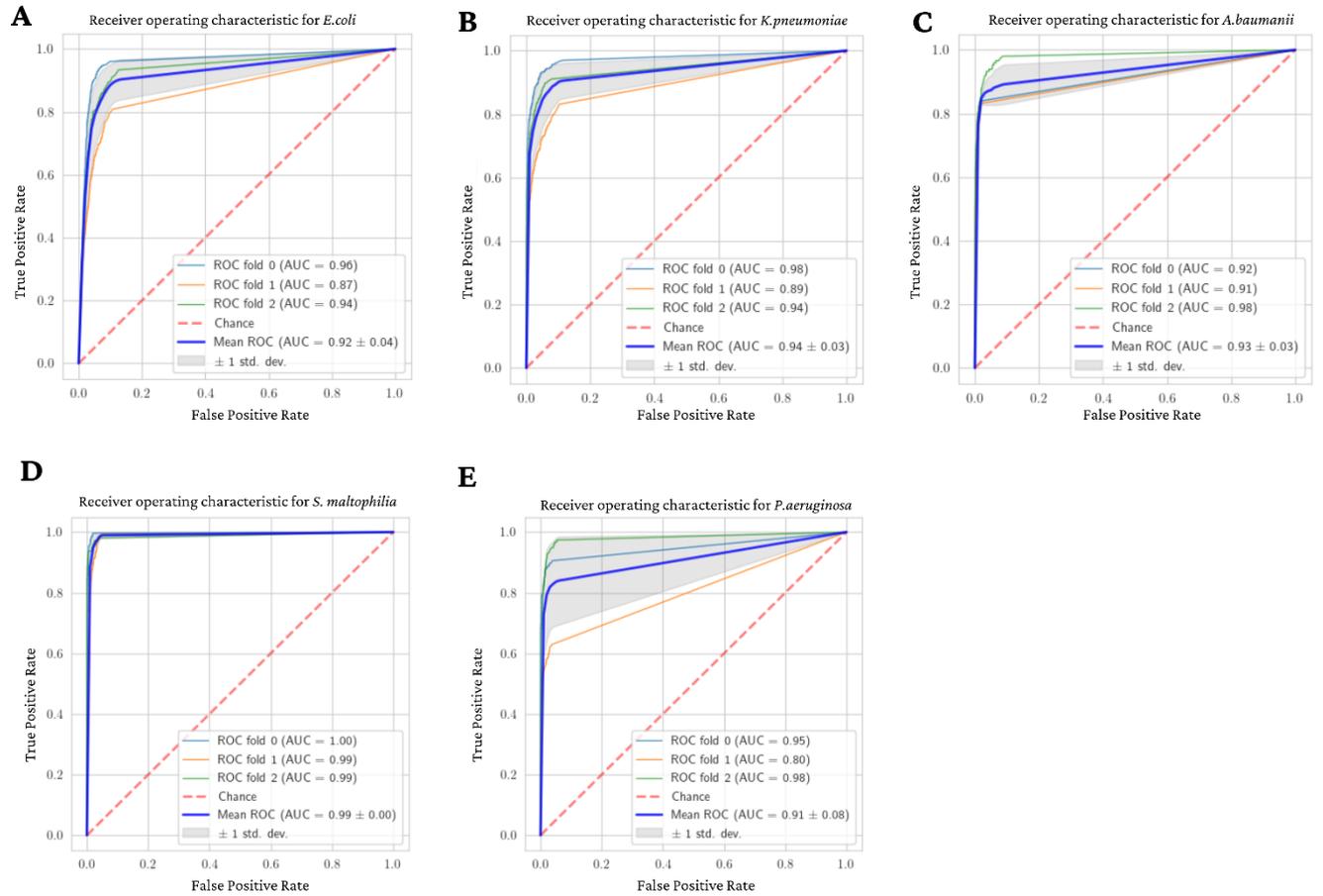

Supplementary Figure 1. ROC AUC scores for the classification of studied Gram-negative bacteria: **A)** *Escherichia coli,* **B)** *Klebsiella pneumoniae*, **C)** *Acinetobacter baumanii,* **D)** *Stenotrophomonas maltophilia*, **E)** *Pseudomonas aeruginosa*

**Appendix B**

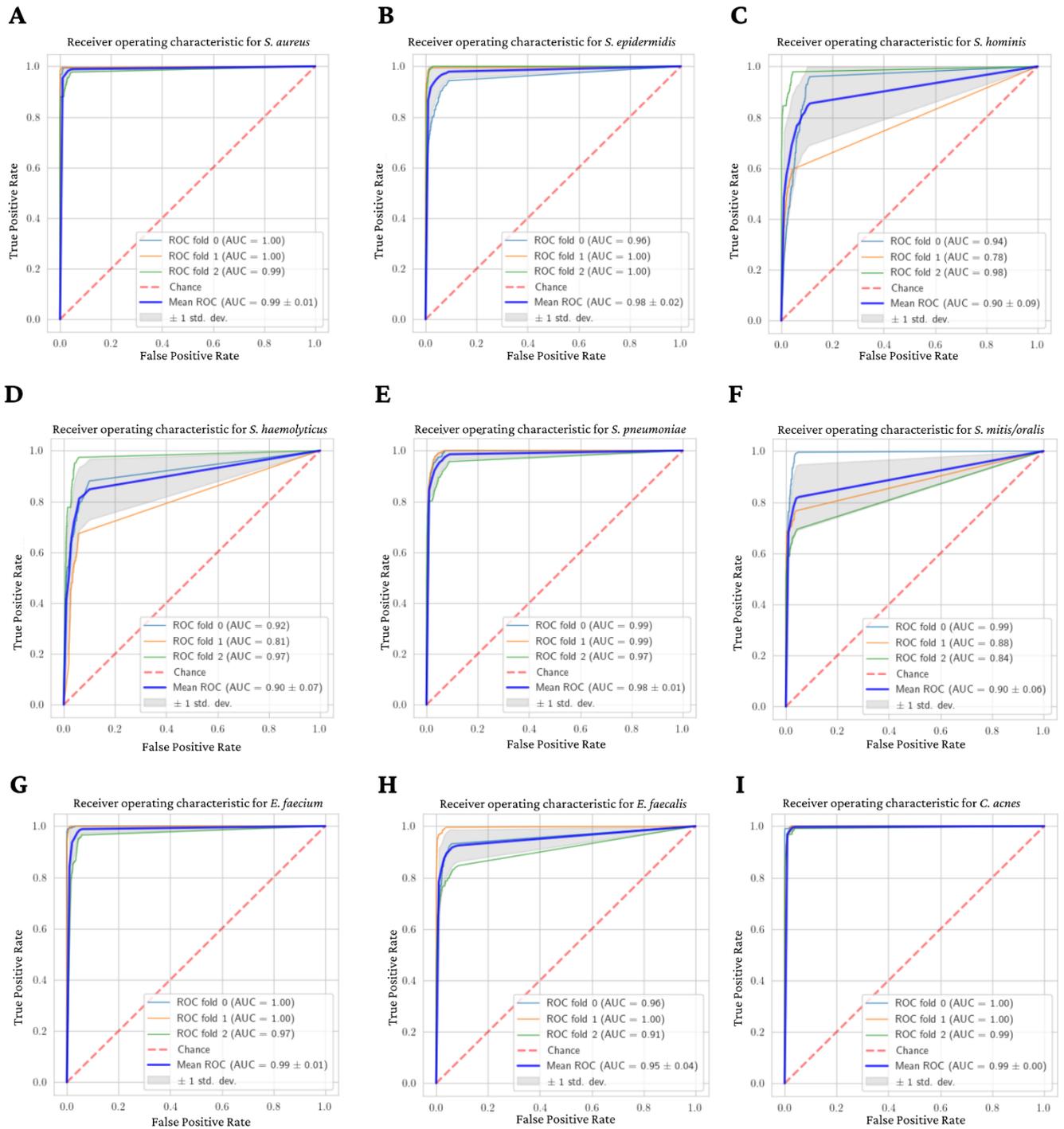

Supplementary Figure 2. ROC AUC scores for the classification of studied Gram-positive bacteria: **A)** *Staphylococcus aureus* **B)** *Staphylococcus epidermidis,* **C)** *Staphylococcus hominis* **D)** *Staphylococcus haemolyticus,* **E)** *Streptococcus pneumoniae,* **F)** *Streptococcus mitis/oralis,* **G)** *Enterococcus faecium,* **H)** *Enterococcus faecalis,* **I)** *Cutibacterium acnes.*

## Appendix C

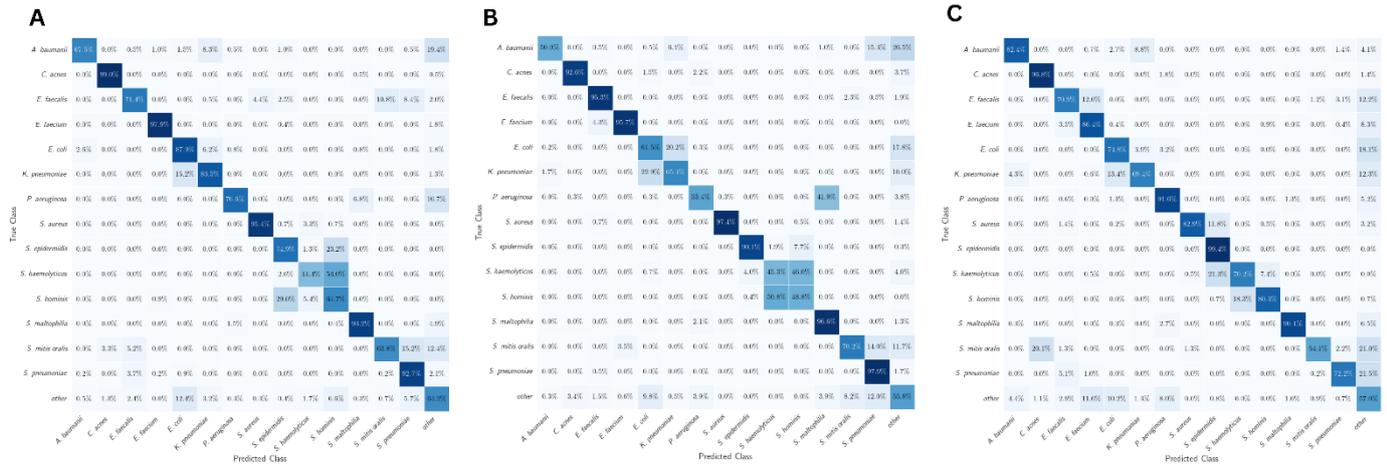

Supplementary Figure 3. Confusion matrix of the tested dataset for each of three – folds (A, B and C), showing the classification accuracy of individual bacterial species in our deep learning model.

## Appendix D

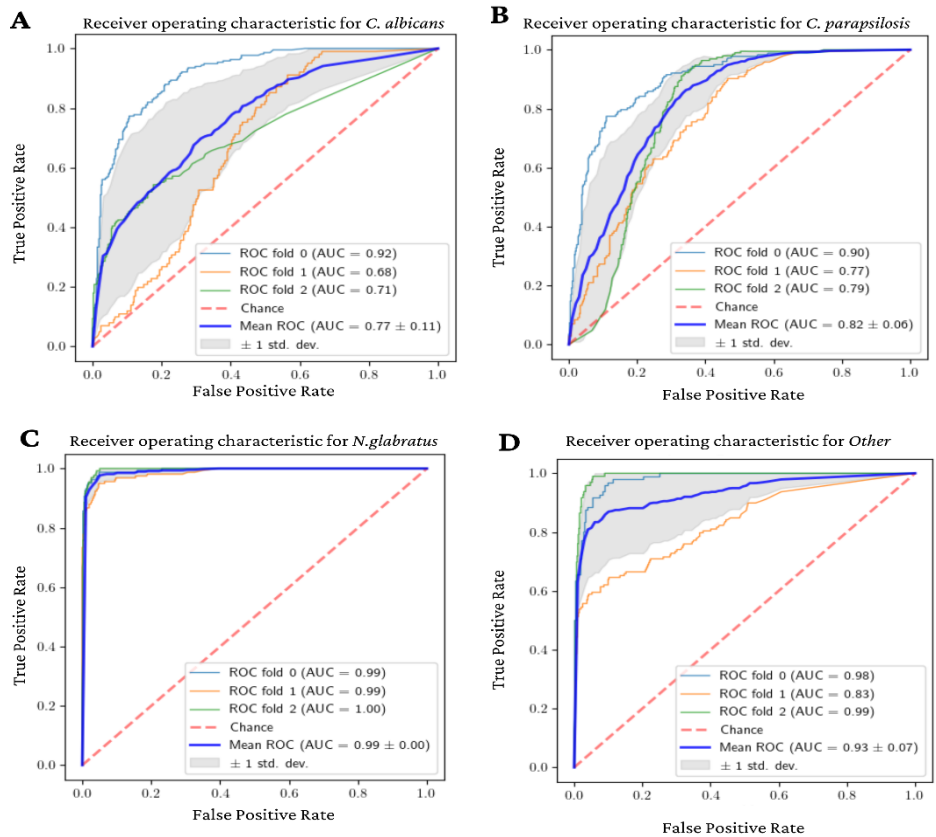

Supplementary Figure 4. ROC AUC scores for the classification of studied fungal species **A)** *C.albicans,* **B)** *C.parapsilosis*, **C)** *N.glabratus,* **D)** Other

**Appendix E**

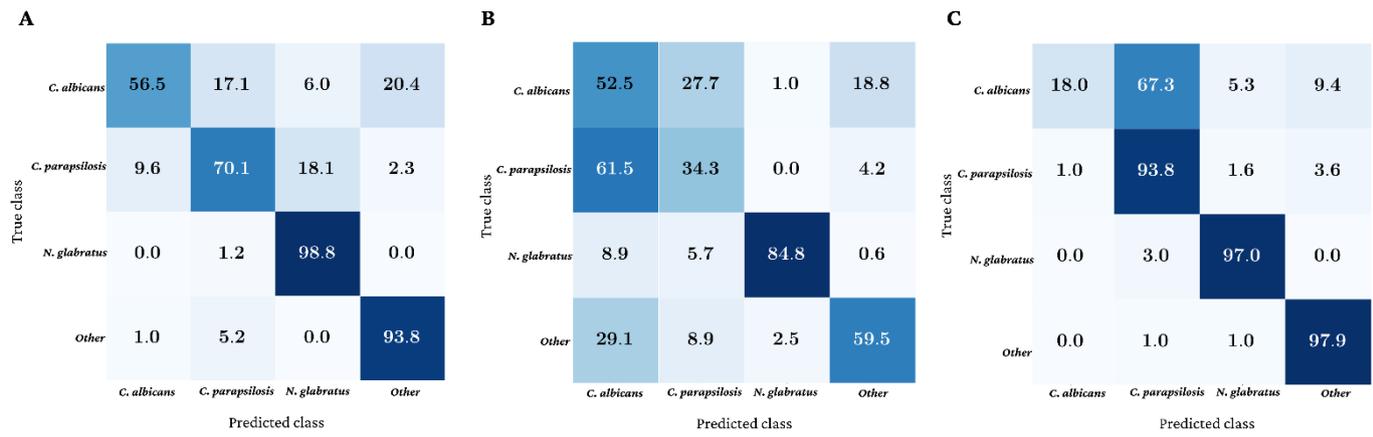

Supplementary Figure 5. Confusion matrix of the tested dataset for each of three – folds (A, B and C), showing the classification accuracy of individual fungal species in our deep learning model.

.